\documentstyle[aps,preprint]{revtex}

\begin{document}

\title{
PSEUDOSCALAR AND SCALAR MESON MASSES AT FINITE TEMPERATURE}
\author{
A. Barducci, M. Modugno, G. Pettini}
\address{Dipartimento di Fisica, Universit\`a di Firenze,
I-50125 Firenze, Italy\\
and Istituto Nazionale di Fisica Nucleare, Sezione di Firenze,
I-50125 Firenze, Italy}
\author{R. Casalbuoni\footnote{On leave from Dipartimento di Fisica,
Universit\`a di Firenze, I-50125 Firenze, Italia}, R. Gatto}
\address{D\'epartement de Physique Th\'eorique, Universit\'e de Gen\`eve,
CH-1211 Gen\`eve 4, Switzerland}

\maketitle
\vspace{0.5cm}
\centerline{UGVA-DPT 1998/09-1015}
\centerline{Firenze Preprint - DFF - 320/09/1998}
\vspace{-0.5cm}

\draft

\begin{abstract}
The composite operator formalism is applied to QCD at finite temperature
to calculate the masses of scalar and pseudoscalar mesons. In particular the
ratio of the sigma mass to the pion mass is an interesting measure of
the degree of chiral symmetry breaking at different temperatures.
We calculate the temperature $T^{*}$ at which $M_{\sigma}(T)\leq2M_{\pi}(T)$,
above which the
sigma partial width  into two pions vanishes. We find
$T^{*}=0.95~T_{c}$ (where $T_c$ is the critical temperature for the chiral
phase transition), within the full effective potential
given by the formalism. We find that  an expansion {\it a-la} Landau of the
effective potential around the critical point in the limit of small quark mass
provides for a very good  determination of $T^{*}$.
\end{abstract}



\section{Introduction}

The concept of chiral symmetry breaking in strong interactions
goes back to the pioneering work by Nambu \cite{nambu}, and it has
been widely explored and discussed since that time. Restoration of
chiral symmetry at high densities was proposed by Lee and Wick \cite{wick}
already in 1974. Theoretical procedures for studying chiral symmetry
restoration
at high temperatures were proposed on the same year by Dolan and
Jackiw \cite{dolan} and by Weinberg \cite{weinberg}.

QCD at finite temperature and density has attracted  much
interest. For latest studies of the QCD phase diagram see for instance
\cite{phase}.

 Physical applications are to high energy heavy ion
collisions \cite{satz}  and to the physics of the early universe \cite{linde}.

A review of lattice calculations can be found for instance in Ref.
\cite{karsch}.
The Nambu-Jona-Lasinio model has been used under different approximations
to obtain indications on the chiral transition
(see for instance \cite{bernard,volkov,vogel,hufner,hatsuda,flork}).

A number of peculiar features may emerge due to the chiral phase transition
and recently much attention has been attracted by the possibility of
disoriented chiral condensate formation (see for instance
\cite{bjorken,rajagopal,blaizot}).

It is strongly suspected that a single
transition occurs rather than separate transitions  for deconfinement and
chiral-symmetry.
We had suggested a heuristic argument indicating that, at least for
zero density, the critical temperature for chiral transition, $T_c$,
 coincides with that for deconfinement, $T_d$ \cite{heuristic}.
The  order
parameters usually considered cover extreme and opposite ranges. The thermally
averaged Polyakov loop is suitable in the limit of infinite quark masses
to describe the transition from the confined to the non-confined phase.
The other extreme is the limit of vanishing quark masses, where the
thermally averaged
quark-antiquark bilinears are the typical order
parameters for chiral symmetry transition.

We shall deal  here with the chiral transition,
concentrating on the thermally averaged quark bilinears
at finite temperatures as order parameters for chirality.
For light massive quarks, such as $u$ and $d$,
rigorously, chiral symmetry  is already
broken in the Lagrangian, but we can still retain
the notion of phase transition, by looking at the region of $T$ where
the condensate has a rapid variation. The current quark mass plays
a role analogous to that of an external magnetic field in the ferromagnetic
transition, as it explicitly violates the chiral symmetry whose restoration
characterizes the phase transition.

The analysis will be based on a composite operator
formalism at finite temperature.
The formalism makes use of an effective action for composite
operators \cite{bcd,masst}. Actually the effective potential admits of
a Landau expansion around the critical point
and thus the behaviour of the condensate is well reconstructed by
knowing the coefficients (which are
infrared safe) and the critical exponents. We shall compare with
the results following from a Landau expansion and show the general agreement.

Within the composite operator formalism we had already discussed the
$T$ dependence of
$f_{\pi}(T)$ \cite{fpit}, in the whole range of temperatures
up to $T_c$.
We shall here mainly deal with the ratio of the scalar to pseudoscalar mass
$M_{\sigma}/M_{\pi}$ at varying temperature. This ratio has a peculiar
theoretical interest as a sensible indicator of the degree of chiral
symmetry breaking.

Let us for the moment neglect the small $u$ and $d$ quark
masses.
 When chiral symmetry is restored the two mesons
are degenerate in mass. The ratio is then equal to unity. At zero temperature
instead in the broken chiral phase the pion is a goldstone and has vanishing
mass, while the sigma has a finite mass from the chiral condensation. Thus one
expects the mass ratio to decrease from $\infty$ to one at the chiral
transition.
Quark masses will change this picture quantitatively leading to a decrease from
a finite value to a value close to one.

Beyond some temperature, before approaching the critical value for the
transition, the sigma will not have phase space left to decay into two pions.
Its instability will decrease when increasing the temperature, and the channel
into two pions will finally be  suppressed at some temperature in the vicinity
but lower than the critical temperature. The decay channel into two photons
will
still be available and become the dominant channel \cite{hatsuda,pisarski}.

It is too early to say whether a possible experimental signature for the
transition
may be inferred from such behaviour. An accurate determination of the mass
ratio
versus temperature is in itself of theoretical interest as it constitutes a
significant parameter for the degree of chiral symmetry breaking at
finite temperature. This we shall do in this note in the composite operator
formalism by making use of parameter values obtained from fits to zero
temperature QCD.

In Section 2 we shall summarize the main results of the application of the
composite operator formalism in QCD. In Section 3 we determine the relevant
QCD
parameters and discuss the hadronic masses at non zero temperature. The
temperature
dependence of the observables is discussed in Section 4, where we also
calculate
the temperature at which the sigma can no longer decay via strong interaction.

\section{Effective action for QCD composite operators}
\label{sec:effact}

Following Ref.\ \cite{bcd} the zero temperature Euclidean
effective action for an $SU(N)$ QCD-like gauge theory is
\begin{equation}
\Gamma( {\bf\Sigma} ) = - {\rm Tr}\ln\left[{\bf S}_0^{-1} + {\delta
\Gamma_2 \over\delta{\bf S}}\right] - {\rm Tr}\left[
{\bf \Sigma}{\bf S}\right] -{\bf \Gamma}_2({\bf S})
+ counterterms
\label{effact}
\end{equation}
where ${\bf S}^{-1}_0= (i\hat{p}- {\bf m})$, {\bf m}
is the bare quark mass matrix and ${\bf \Gamma}_2({\bf S})$
is the sum of all the
two-particle irreducible vacuum diagrams with fermionic propagator
${\bf S}$ and ${\bf \Sigma} = -\delta{\bf \Gamma}_2/\delta{\bf S}$.
At two-loop level ${\bf \Gamma}_2={1\over2}{\rm Tr}({\bf S}\Delta{\bf S})$,
where $\Delta$ is the gauge boson propagator, so that
${\bf \Sigma} = - \Delta{\bf S}$,
${\rm Tr}\left[{\bf S}~{\delta{\bf \Gamma}_2/\delta{\bf S}}\right] =
2{\bf \Gamma}_2$
and one can rewrite Eq.\ (\ref{effact}) in terms of ${\bf \Sigma}$
\begin{eqnarray}
\Gamma( {\bf\Sigma} ) &=& - {\rm Tr}\ln\left[{\bf S}_0^{-1} -{\bf \Sigma}
\right] + {\bf \Gamma}_2({\bf \Sigma}) + counterterms\nonumber\\
&=&- {\rm Tr}\ln\left[{\bf S}_0^{-1} -{\bf \Sigma}
\right] + {1\over 2}{\rm Tr}\left({\bf \Sigma}\Delta^{-1}{\bf \Sigma}
\right) + counterterms
\label{effact1}
\end{eqnarray}

Here the variable ${\bf \Sigma}$ plays the role of a dynamical variable.
At the minimum of the functional action, that is when the Schwinger-Dyson
equation is satisfied, ${\bf \Sigma}$ is nothing but the fermion self-energy.
A parametrization for ${\bf \Sigma}$, employed in \cite{bcd} was
\begin{equation}
 {\bf\Sigma}=({\bf s} + i \gamma_5{\bf p})f(k)\equiv {\bf\Sigma}_s
+i\gamma_5 {\bf\Sigma}_p
\label{selfen}
\end{equation}
with a suitable Ansatz for $f(k)$, and with {\bf s} and {\bf p}
scalar and pseudoscalar constant fields respectively
to be taken as the variational parameters.

The effective potential one obtains from Eq.\ (\ref{effact1})
(see Ref. \cite{bcd}) is
\begin{eqnarray}
V={\Gamma\over\Omega}&=& -{8\pi^2 N\over 3 C_2 g^2}
\int{d^4 k\over(2\pi)^4}{\rm tr}\left[{\bf \Sigma}_s\Box_k{\bf \Sigma}_s
+{\bf \Sigma}_p\Box_k{\bf \Sigma}_p\right]-\nonumber\\
&&- N~{\rm Tr}\ln\left[i\hat{k} - \left({\bf m}+{\bf \Sigma}_s
\right)-i\gamma_5{\bf \Sigma}_p\right] + \delta Z~{\rm tr}({\bf m}~{\bf s})
\label{effpot}
\end{eqnarray}
where $C_2$ is the quadratic Casimir of the fermion representation
(for $SU(3)_{c}$  $C_{2}=4/3$). Furthermore ${\bf \Sigma}_s \equiv
\lambda_{\alpha} s_{\alpha} f(k)/ \sqrt{2}$,
${\bf \Sigma}_p \equiv \lambda_{\alpha} p_{\alpha} f(k)/ \sqrt{2}$,
${\bf m} \equiv \lambda_{\alpha}m_{\alpha}/ \sqrt{2}$ ($\alpha=0,\cdots,8$,
$\lambda_0=\sqrt{2/3}~I$, $\lambda_i =$ Gell-Mann matrices,
$i=1,\cdots,8$), $g$ is the gauge coupling constant and $\Omega$ is the
four-dimensional volume.
In Eq.\ (\ref{effpot}) $\delta Z$  has a divergent part to
compensate the leading divergence
proportional to  ${\rm Tr}({\bf m}~{\bf s})$ in the logarithmic term.
We remark that both the current mass {\bf m} and the
self-energy ${\bf \Sigma}$ are, in general, matrices in flavour space.
However, as discussed in Refs.\ \cite{bcd,masst} if we neglect the mixing
between different flavours originating, for instance, from terms such as
the 't Hooft determinant \cite{hooft}, it follows that only the flavour
diagonal elements of the fermion self-energy and mass can be different
from zero at the minimum. With vanishing off-diagonal terms, the effective
potential decomposes into the sum of $n_{f}$ contributions ($n_{f}$~=~number
of flavours), one for each flavour. Therefore, to study the minima of the
effective potential, it is formally sufficient to consider a single
contribution. Of course, the choice of a given flavour number will reflect
in the particular parameters assumed. In the present paper, as in Ref.\
\cite {bcd}, we will take $n_{f}=3$ and a number of colour $N=3$.
The value of the parameters will be specified later on.

As far as the choice for the function $f(k)$ is concerned, in QCD, the
operator product expansion suggests (neglecting logarithmic corrections)
to take for $f(k)$ a momentum behaviour as $1/k^{2}$ for large $k^2$.
We have chosen as a variational Ansatz \cite{giulio1}
\begin{equation}
f(k)={M^{3}\over
 k^2+M^2}
\label{fope}
\end{equation}
where $M$ is a momentum scale which is expected to be of the order
of $\Lambda_{QCD}$.

To extend the zero-temperature theory to finite temperature we can still
work with continuous energies by substituting for the sum over discrete
energies $\omega_{n}=(2n+1)\pi/\beta$ (where $\beta=1/T$)
a sum of integrals over continuous energies by means of
the Poisson's formula \cite{stak}
\begin{eqnarray}
\int{d^{4}k\over (2\pi)^{4}}~f(k)&\rightarrow & {1\over\beta}\sum_{n=-\infty}
^{n=+\infty}\int{d^{3}k\over (2\pi)^{3}}~f(\omega_{n},{\bf k})\nonumber\\
&=&\sum_{n=-\infty}^{n=+\infty}~(-)^{n}~\int{d^{4}k\over (2\pi)^{4}}~
f(k)~e^{{\displaystyle{in\beta k_{0}}}}
\label{poisson}
\end{eqnarray}

This substitution corresponds to the imaginary-time formalism after using
the Poisson's formula and allows for an equivalent version af the
Dolan-Jackiw finite temperature Feynman rules \cite{dolan,weinberg}.

In conclusion the final form for the effective potential, for a quark of mass
$m$, is (see Ref. \cite{giulio1})
\begin{eqnarray}
V={\Gamma\over\Omega}&=& -{8\pi^2 N\over 3 C_2 g^2(T)}
\int{d^4 k\over(2\pi)^4}\left[\Sigma_{s}\Box_k \Sigma_{s}
+ \Sigma_p\Box_k \Sigma_p\right]-\nonumber\\
&&- 2N~\sum_{n=-\infty}^{n=+\infty}~(-)^{n}~\int{d^{4}k\over (2\pi)^{4}}~
\ln\left[k^2 +\left(m + \Sigma_s\right)^2+\Sigma^{2}_{p}\right] +
\delta Z~ms
\label{effpot1}
\end{eqnarray}
where $\Sigma_{s}\equiv s~f(k)$ and $\Sigma_{p}\equiv p~f(k)$.

Let us now comment on the choice for the gauge coupling constant.
As suggested by asymptotic freedom and renormalization-group
considerations, we expect the strong forces to weaken at high
temperature \cite{perry}. We shall then assume that in the UV region
the coupling constant $g(T)$ depends logarithmically on the
temperature $T$. We take into account this assumption by writing
\begin{eqnarray}
{g^{2}(T)\over 2\pi^2}\equiv {1\over c(T)}&\equiv&
{1\over c_{0}+c_{1}(T)}\nonumber\\
&=&{1\over {\displaystyle{c_{0}+{\pi^2\over b}\ln\left(1
+\xi~{T^2\over M^2}\right)}}}
\label{coupling}
\end{eqnarray}
where $b=24\pi^2/(11N-2n_{f})$ and we will discuss the parameters
$c_{0}$, $M$ and $\xi$ later on.

At $T\neq 0$ the effective potential does not acquire any
extra divergence with respect to the $T=0$ case.
The renormalization at $T=0$ can be performed by adding a counterterm
and requiring that the derivative of the effective potential
with respect to the term which breaks explicitly the chiral
symmetry, evaluated at the minimum, satisfies for each flavour the
renormalization condition \cite{bcd}
\begin{equation}
\lim_{m\rightarrow 0}~\left.{\delta V\over \delta\left(m
\langle{\bar\psi}\psi\rangle_{0}\right)}\right|_{\rm min}~=~1
\label{normcond}
\end{equation}
The fermion condensate $\langle{\bar\psi}\psi\rangle_{T}$
is related to the minimum ${\bar s}(T)$ of the effective potential,
renormalized at the scale $M$ through the relation (see Ref.\ \cite{bcd})
\begin{equation}
\langle{\bar\psi}\psi\rangle_{T}={3M^{3}\over g^2(T)}{\bar s}(T)
\label{condenst}
\end{equation}
and $\langle{\bar\psi}\psi\rangle_{0}\equiv
\langle{\bar\psi}\psi\rangle_{T=0}$.

At finite $T$ we do not have any additional divergence with respect
to the $T=0$ case; nevertheless, in order to satisfy the generalization
of the normalization condition (\ref{normcond}) at finite $T$ we have to add
a finite counterterm to that determined at $T=0$ \cite{masstmu}.
Finally we recall that with the appropriate normalization for the pion field
(see next Section), Eq.\ (\ref{normcond}) and its generalization at finite $T$
is also equivalent to the Adler-Dashen formula
\begin{equation}
M^{2}_{\pi}(T)~f^2_{\pi}(T)~=~-2m~\langle{\bar\psi}\psi\rangle_{T}
\label{adash}.
\end{equation}

\section{Hadronic masses at $T\neq 0$ and determination of the parameters}
\label{sec:hadmass}
To compute the masses of the scalar and pseudoscalar mesons ($\sigma$
and $\pi$ respectively) one has to take the second derivative of the
effective potential (\ref{effpot1}) with respect to the scalar
field $s$ and the pseudoscalar field $p$, evaluated at the minimum
(for the sake of simplicity we are
assuming that the up and down quarks are degenerate in mass and
$m=(m_{u}+m_{d})/2$ in Eq.\ (\ref{effpot1})).

The actual values of the masses will be obtained by multiplying the second
derivative by the appropriate factor $a$ that relates the physical
fields $\varphi_{\pi}$ ($\varphi_{\sigma}$) to $p$ ($s$) according to
$\varphi_{\pi}~=~a~p,~\varphi_{\sigma}~=~a~s$.
This factor can be obtained in terms of the decay constant $f_{\pi}$
through standard arguments of current algebra \cite{bcd}. One gets
\begin{eqnarray}
a~&=&~-~{f_{\pi}\over {\sqrt{2}}{\bar s}}\nonumber\\
M^{2}_{\sigma}~&=&~{\partial^2 V\over \partial \varphi^2_{\sigma}}\Big|
_{\rm min}~=~{1\over a^2}~{\partial^2 V\over \partial s^2}\Big|_{\rm min}
~=~{2{\bar s}^2\over f^2_{\pi}}~
{\partial^2 V\over \partial s^2}\Big|_{\rm min}\label{masses}\\
M^{2}_{\pi}~&=&~{\partial^2 V\over \partial \varphi^2_{\pi}}\Big|
_{\rm min}~=~{1\over a^2}~{\partial^2 V\over \partial p^2}\Big|_{\rm min}
~=~{2{\bar s}^2\over f^2_{\pi}}~
{\partial^2 V\over \partial p^2}\Big|_{\rm min}\nonumber
\end{eqnarray}
where ${\bar s}$ is the extremum of the effective potential in the presence
of a bare mass.

To derive a more physically transparent expression for the masses
of the mesons as expressed by Eq.\ (\ref{masses})
through the second derivative of the effective potential, we will
use the gap equation and the generalization at finite $T$ of
the normalization condition (\ref{normcond}).
The extremum condition is
\begin{eqnarray}
{\partial V\over \partial s}=0 ~\rightarrow ~{N\over {\bar s}}&&\left[
-2 c(T) \int{d^4 k\over (2\pi)^4}{\bar \Sigma}_{s}(k)\Box_k
{\bar \Sigma}_{s} (k)\right.\nonumber\\
&&\left.-4\sum_{n} (-)^{n}\int{d^4 k\over (2\pi)^4}
{\left( m+{\bar \Sigma}_{s} (k)\right){\bar \Sigma}_{s}(k)\over
\left[k^2+\left(m+{\bar \Sigma}_{s}(k)\right)^2\right]}
~e^{\displaystyle{in\beta k_{0}}}\right]+~m~\delta Z=0
\label{extrcond}\\
{\partial V\over \partial p}=0 \rightarrow  {\bar p}=0
\label{extrcond1}
\end{eqnarray}
where ${\bar \Sigma}_{s}(k)\equiv{\bar s}~f(k)$.

As far as the normalization condition is concerned, by using the
relation (\ref{condenst}) between the scalar field at the minimum
and the scalar condensate we shall write in general
\begin{equation}
\delta Z=N\left[{M^3\over 2\pi^2}c(T)+{4\over s_{0}}\sum_{n}
(-)^{n}\int{d^{4}k\over (2\pi)^{4}}{\Sigma_{0}(k)\over
k^2+\Sigma_{0}^{2}(k)}~e^{\displaystyle{in\beta k_{0}}}\right]
\label{deltaz}
\end{equation}
with $\Sigma_{0}(k)=s_{0}f(k)$ and where $s_{0}$ is the minimum
of the effective potential in the massless case.

Now, by using the gap equations (\ref{extrcond})-(\ref{extrcond1})
and the normalization
condition (\ref{deltaz}) in the explicit expressions (\ref{masses})
for the meson masses, we can eliminate $\delta Z$ and $c(T)$ to finally
obtain, for a quark of mass $m$:
\begin{equation}
M^{2}_{\pi}=-2{m\over f^{2}_{\pi}}\langle{\bar\psi}\psi\rangle_T+{8Nm\over
f^{2}_{\pi}}\sum_{n}(-)^{n}\int{d^{4}k\over (2\pi)^{4}}
\left[{{\bar\Sigma}_{s}\over k^2+\left(m+{\bar\Sigma}_{s}\right)^2}-
{{\bar s}\over s_{0}}~{\Sigma_{0}\over k^2+\Sigma_{0}^2}\right]
~e^{\displaystyle{in\beta k_{0}}}
\label{emmepai2}
\end{equation}
\begin{equation}
M^{2}_{\sigma}=M^{2}_{\pi}+{16N\over f^{2}_{\pi}}
\sum_{n}(-)^{n}\int{d^{4}k\over (2\pi)^{4}}
{\left(m+{\bar\Sigma}_{s}\right)^{2}{\bar\Sigma}^{2}_{s}
\over \left[k^2+\left(m+{\bar\Sigma}_{s}\right)^2\right]^{2}}
~e^{\displaystyle{in\beta k_{0}}}
\label{emmesig2}
\end{equation}

We notice that Eq.\ (\ref{emmepai2}) reproduces, in the small
mass limit, the Adler-Dashen formula (\ref{adash}) and differs
from this formula only by terms of order $m^2$.

In order to fix the values of the parameters $c_0$ and $s_{0}(T=0)$
we follow the same procedure of Ref.\ \cite{bcd}. In the massive case
the effective potential is UV divergent and from the normalization
condition (in the small-mass limit) and gap equation one is able
to fix, through a self-consistency relation, the values $c_{0}=0.554$.
By inserting this value in the gap equation for massless quarks, one finds
that chiral symmetry is spontaneously broken  (at $T=0$) via a minimum
of $V$ located at $s_{0}(T=0)=-4.06$ and the point $s_{0}(T=0)=0$ is a local
maximum. To determine the mass scale $M$ and the quark masses from
the experimental data, one has to derive the explicit expressions for the
masses and for the decay constants of the pseudoscalar octet mesons which
are the pseudo-Goldstone bosons of chiral symmetry breaking.
These expressions constitute a system of coupled equations which
we have solved by an approximation method.
The experimental inputs are the decay constant and mass of the
charged pion, the charged kaon mass and the electromagnetic
mass difference between the neutral and the charged kaon. The
outputs of the numerical fit for the octet meson masses (agreement
within 3\%) are the masses of the $u$, $d$ and $s$ quarks and the mass
parameter $M$. The values we get are the following
\cite{bcd,masst,tokyo}
\begin{eqnarray}
M~&=&~280~MeV\nonumber\\
m_{u}~&=&~8~MeV\nonumber\\
m_{d}~&=&~11~MeV\label{qmasval}\\
m_{s}~&=&~181~MeV\nonumber\\
m~&\equiv&~{m_{u}+m_{d}\over 2}=9.5~MeV\nonumber
\end{eqnarray}
Finally, the parameter $\xi$, which appears in the expression
(\ref{coupling}) for the running coupling constant, has to be determined
on phenomenological grounds. If we compare our model in the low-$T$
regime with the results of Ref.\ \cite{leut} we find
$\xi=0.44 (n^{2}_{f}-1)/n_{f}$ for $n_{f}$ flavours, which gives
$\xi\simeq 1$ for $n_{f}=3$ \cite{fpit}.

With these values for the relevant parameters it is easy to
determine the masses of the pseudoscalar and scalar mesons at $T=0$
(we are using the value $f_{\pi}=93~MeV$ for the pion decay constant)
\begin{eqnarray}
M_{\pi}\simeq 135~MeV,\qquad
M_{\sigma}\simeq 630~MeV
\label{masspisi}
\end{eqnarray}
and (again at $T=0$) the fermionic condensates at $M$ for massless
and massive quarks
\begin{eqnarray}
\langle{\bar\psi}\psi\rangle_0&\simeq&(-197~MeV)^3\qquad(massless\quad
quarks)\\
\langle{\bar\psi}\psi\rangle_0&\simeq&(-200~MeV)^3\qquad(massive\quad quarks).
\label{conds}
\end{eqnarray}
The value we have obtained for the mass of the scalar meson,
$M_{\sigma}\simeq 630~MeV$, accords very well with those
obtained by other authors within different approaches
to the study of dynamical chiral symmetry breaking in QCD
\cite{hatsuda,flork}.

\section{The scalar-pseudoscalar $M_{\sigma}(T)/M_{\pi}(T)$ mass ratio}
\label{sec:masratio}
It is well known that as the temperature $T$ increases we expect
some changes in the hadronic properties of matter as a consequence of
a possible phase transition.

For instance, a physical quantity as $f_{\pi}(T)$ will
decrease as $T$ increases \cite{fpit}. This means that the pion tends
to decouple from matter (quarks and leptons) as $T$ increases.
As far as the pion mass is concerned, $M_{\pi}(T)$ has a weak
dependence on the temperature. In fact the value of $M_{\pi}(T)$ is
dominated by the current quark mass for temperatures below the critical
value, whereas it becomes independent of $m$ at the critical temperature
or above it. This should be expected because, for $T\simeq T_{c}$, the
pion becomes an ordinary resonance and its mass is dominated by
$\Lambda_{QCD}$ and not by $m$. This is a clear sign that the pion loses
its Goldstone nature once the approximate chiral symmetry is restored
\cite{masst}.

The scalar meson mass $M_{\sigma}(T)$ will decrease in association
with partial chiral symmetry restoration in hot matter. In fact
the scalar $\sigma$ meson can be described as a $q{\bar q}$ quasi
bound state and its mass is about twice the constituent quark mass
\cite{hatsuda,bcd,scadron}.
Thus the $\sigma$-mass will decrease substantially with increasing
temperature and this behaviour is mainly determined by the $T$-dependence
of the constituent quark mass. The decreasing of the
$\sigma$-mass is expected from the restoration of chiral symmetry at
$T=T_{c}$. In fact $M_{\sigma}\rightarrow 0$ for $m=0$ at $T=T_{c}$.
As a consequence $M_{\sigma}(T)$ will become smaller than $2M_{\pi}(T)$
at some temperature $T^{*}<T_{c}$ and the
$\sigma$-meson, which has a large width at $T=0$ due to the decay
$\sigma\rightarrow 2\pi$, would appear as a sharp resonance at high
temperature since the amplitude for $\sigma\rightarrow 2\pi$
would vanish above $T^{*}$ (near the critical
point of the chiral phase transition).
Thus one can expect to have a better chance of seeing the $\sigma$-meson
emerge as a clear, narrow state at non zero temperature.
In fact, since for $T>T^{*}$ the decay width of $\sigma$ by strong
interaction vanishes, only the electromagnetic width coming from
$\sigma\rightarrow 2\gamma$ is left \cite{hatsuda,pisarski}.

Within our model we are able to make some rough estimate of the
temperature $T^{*}$ at which this phenomenon takes place by simply
studying the behaviour of the meson masses (\ref{emmepai2}),
(\ref{emmesig2}) as a function of the temperature.
However, since we are, as already said, mainly interested in determining
the temperature $T^{*}$ at which $M_{\sigma}(T)\leq2M_{\pi}(T)$,
instead of evaluating separately $M_{\sigma}(T)$ and $M_{\pi}(T)$
it will be sufficient to study the ratio $M^{2}_{\sigma}/M^{2}_{\pi}$
\begin{equation}
{M^{2}_{\sigma}\over M^{2}_{\pi}}~=~1~-~{24\over m}~
{\displaystyle{\sum_{n}(-)^{n}\int {d^{4}k\over (2\pi)^{4}}
{\left(m+{\bar \Sigma}_{s}\right)^{2}\over \left[ k^{2}
+\left(m+{\bar \Sigma}_{s}\right)^{2}\right]^{2}}~{\bar \Sigma}_{s}^{2}
~e^{\displaystyle{in\beta k_{0}}}}
\over\displaystyle{
\langle{\bar\psi}\psi\rangle_T~-12~
{\sum_{n}(-)^{n}\int {d^{4}k\over (2\pi)^{4}}
\left[{{\bar \Sigma}_{s}\over k^{2}+\left(m+{\bar \Sigma}_{s}\right)^{2}}
~-~{{\bar s}\over  s_{0}}~
{\Sigma_{0s}\over k^{2}+\Sigma^2_{0s}}\right]}~
e^{\displaystyle{in\beta k_{0}}}}}
\label{ratiosq}
\end{equation}
By straightforward calculations at the leading order in $\alpha=m/M$
we obtain
\begin{equation}
{M^{2}_{\sigma}\over M^{2}_{\pi}}~\simeq~1~-~16\pi^2~{s_{0}^{3}(T)\over
c(T)}~{M\over m}~\sum_{n}(-)^{n}\int {d^{4}y\over (2\pi)^{4}}
~{e^{\displaystyle{in\beta M y_{0}}}\over\left[y^{2}\left(y^{2}+1\right)^{2}
+s_{0}^{2}\right]^{2}}
\label{ratiosq1}
\end{equation}
Furthermore, as we expect that the temperature $T^{*}$ will be not far from
the critical temperature of the chiral phase transition, we can start
our investigation by performing a mean-field expansion
{\it a-la} Landau of the effective potential in the small quark-mass limit
around the critical point. Following Ref.\ \cite{masst} we have
\begin{equation}
V={3M^{4}\over 4\pi^{2}}~\left[a_{2}(T)~\varphi^{2}~+~a_{4}(T)~\varphi^{4}
~+~b_{1}(T)~\alpha~s~+~....\right]
\label{landau}
\end{equation}
where $\varphi^2=s^2+p^2$.

If we now compute the masses of the scalar and pseudoscalar mesons
from the expression (\ref{landau})
of the effective potential we get
\begin{equation}
{M^{2}_{\sigma}\over M^{2}_{\pi}}~\simeq~1~-~8{M\over m}~{a_{4}(T)\over
b_{1}(T)}~s_{0}^{3}(T)~+~...
\label{landmasratio}
\end{equation}
By noticing that the generalization at finite $T$ of
the normalization condition (\ref{normcond})
on the effective potential reads now
\begin{equation}
b_{1}(T)~=~2~c(T)
\label{lanormcon}
\end{equation}
and that the coefficient $a_{4}(T)$ is given by the expression
(see Ref.\ \cite{fpit})
\begin{equation}
a_{4}(T)~=~4\pi^2~\sum_{n}(-)^{n}\int {d^{4}y\over (2\pi)^{4}}
{1\over\left[y^{2}\left(y^{2}+1\right)^{2}+s_{0}^{2}\right]^{2}}
~e^{\displaystyle{in\beta M y_{0}}}
\label{aquattro}
\end{equation}
it is easy to check that Eq.\ (\ref{landmasratio})
is nothing but Eq.\ (\ref{ratiosq1}).

Just to have a rough idea of the value of the temperature $T^{*}$
we can push our approximation further by evaluating $a_{4}(T)$ and
$b_{1}(T)$ at $T_{c}$ and using for $s_{0}(T)$ the expression
for the massless minimum around $T_{c}$ \cite{masst}
\begin{equation}
s_{0}(T)~=~a_{s}(T_{c})\left(1-{T\over T_{c}}\right)^{1/2}
~~~~~~~~~~~~~~~~~~~~~~~~~~~T\leq T_{c}
\label{essezerot}
\end{equation}
in Eq.\ (\ref{landmasratio}).

The numerical values for the coefficients relevant for
the evaluation of the mass ratio are
\begin{equation}
a_{4}(T_{c})=0.0033~,~~~~~~~b_{1}(T_{c})=1.25~,~~~~~~~a_{s}(T_{c})=-7.28
\label{coefval}
\end{equation}
and $T_{c}\simeq103 MeV$ \cite{masst}.

By now imposing the condition $M_{\sigma}/2M_{\pi}\leq 1$ we have
a constraint on the temperature ratio $T/T_{c}$
\begin{equation}
{M_{\sigma}\over 2M_{\pi}}\leq 1 ~~~~~~\Longrightarrow~~~~~~{T
\over T_{c}}\geq 0.95
\label{constraint}
\end{equation}
and this constraint fixes $T^{*}=0.95~T_{c}\simeq 98 MeV$.

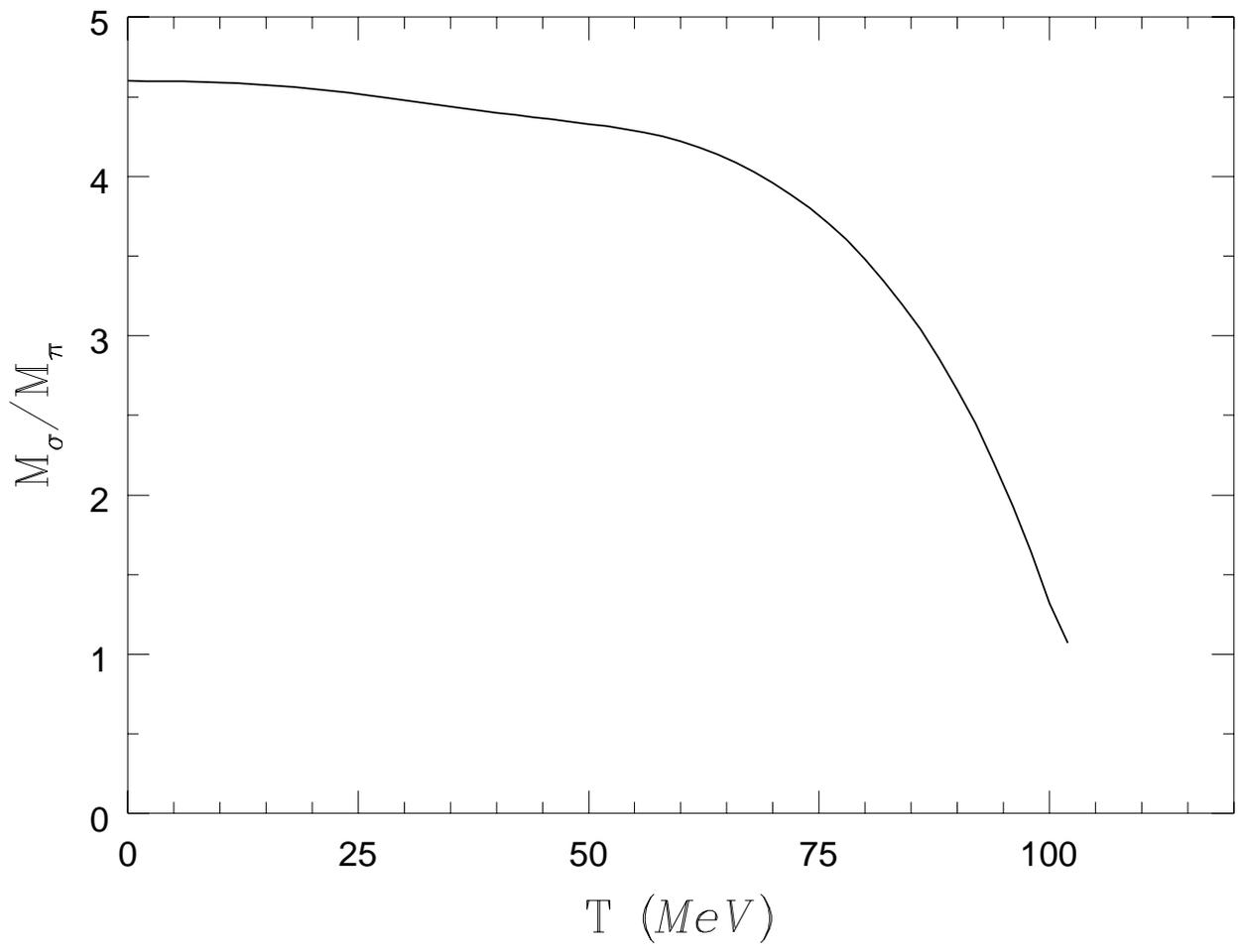
\begin{figure}
\caption{Plot of $M_{\sigma}/M_{\pi}$ versus
$T$ at the leading order in $m$, for $m=9.5~MeV$.
\label{fig1}}
\end{figure}

The approximations we have made up to now (we have evaluated $a_{4}(T)$
and $b_{1}(T)$ at $T_{c}$, and we have chosen for $s_{0}(T)$ the
behaviour as given by Eq.\ (\ref{essezerot})) allow us to study
$M_{\sigma}/M_{\pi}$ only for $T$ very close to $T_{c}$.
To study the mass ratio $M_{\sigma}/M_{\pi}$ as a function of $T$
for all the values of the temperature from
$T=0$ up to $T$ very close to $T_{c}$ we have to go back to Eq.\
(\ref{ratiosq1}) which is valid for any $T$ at the leading order in the
quark mass $m$.
To evaluate $M^2_{\sigma}/M^2_{\pi}$ we can perform the integration
in $y_{0}$ by using the Cauchy theorem and summing up the series;
then, through a numerical calculation we can finally perform
the integration left. In Fig.\ \ref{fig1} we plot the mass ratio
$M_{\sigma}/M_{\pi}$ as a function of $T$ for a mass
$m=(m_{u}+m_{d})/2=9.5~MeV$; from the numerical analysis we can
evaluate again the temperature $T^{*}$
\begin{equation}
{M_{\sigma}\over 2M_{\pi}}\leq 1~~~~~~\Longrightarrow ~~~~~~
{T\over T_{c}}\geq 0.93
\label{resfig1}
\end{equation}
and this constraint now fixes $T^{*}=0.93~T_{c}\simeq 95~MeV$.

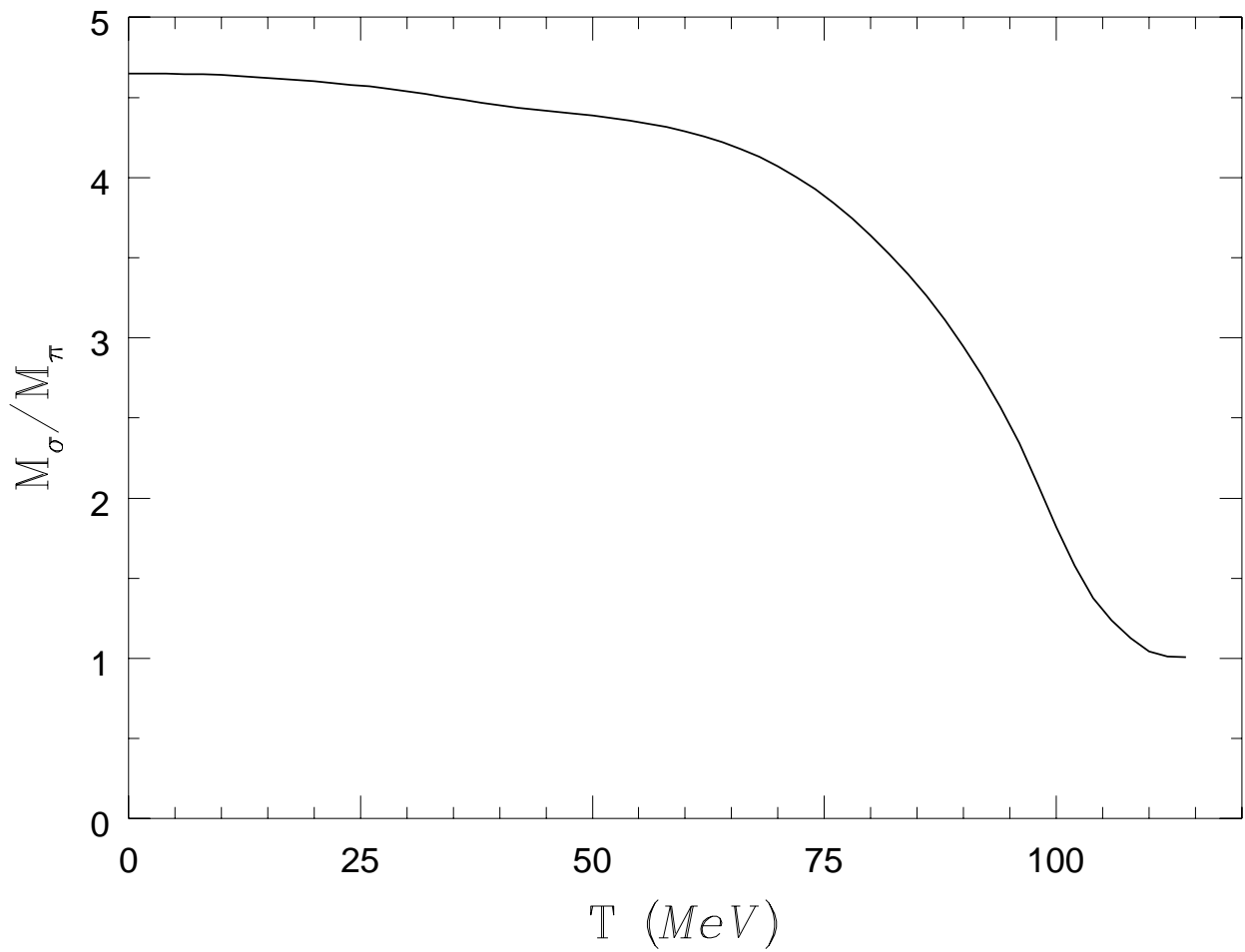
\begin{figure}
\caption{Plot of $M_{\sigma}/M_{\pi}$ versus
$T$ at all orders in $m$, for $m=9.5~MeV$.
\label{fig2}}
\end{figure}

We remark that this result for $T^{*}$ is in complete agreement
with that previously derived by studying the Landau expansion of the
effective potential for $T$ very close to $T_{c}$.
Finally, if one wants to study $M_{\sigma}/ M_{\pi}$ including all
the orders in the quark mass $m$ and for any temperature it is necessary
to go back to the general formula (\ref{ratiosq}).
In Fig.\ \ref{fig2} it is plotted the behaviour
of the mass ratio as a function of $T$ as derived by the explicit
evaluation of Eq.\ (\ref{ratiosq}). In this case the
temperature $T^{*}$ we are looking for is determined again by the
condition
\begin{equation}
{M_{\sigma}\over 2M_{\pi}}\leq 1~~~~~~\Longrightarrow ~~~~~~
{T\over T_{c}}\geq 0.95
\label{resfig2}
\end{equation}
which fixes $T^{*}=0.95~T_{c}\simeq 98~MeV$.
Curiously this result exactly coincides with that obtained
through the Landau expansion and the evaluation of the coefficients
$a_{4}(T)$, $b_{1}(T)$ at $T_{c}$ and by putting for $s_{0}$
the expression (\ref{essezerot}) as suggested by the mean-field theory.

\section{Conclusions}
\label{sec:Conclusions}

We have applied the composite operator formalism to QCD at finite temperature.
After carrying out consistently the renormalization in the effective potential,
we have calculated the masses of the scalar and pseudoscalar mesons.
The relevant QCD parameters have been derived from the expressions for the
masses and
decay constants of the pseudoscalar multiplet, by inserting as inputs
the mass and decay constant of the pion, the charged kaon mass and the
electromagnetic
mass difference between neutral and charged kaon.

A main purpose of the calculation was
to determine the temperature $T^{*}$ at which $M_{\sigma}(T)\leq2M_{\pi}(T)$.
Such temperature will be reached before the critical point, since the
scalar meson
mass $M_{\sigma}(T)$ is expected to decrease
along with partial chiral symmetry restoration in hot matter, and at the
same time
the pion will lose its Goldstone nature.The temperature $T^{*}$ is also
expected not to
lie far from the critical temperature of the chiral transition.
In principle
the $\sigma$-meson would appear as a narrow state at high enough temperature.
 For $T>T^{*}$ its partial width  by strong
interaction has to vanish.

The result of our calculation gives $T^{*}=0.95~T_{c}$, in an
approximation including current quark masses and with the full effective
potential
given by the formalism. We find that this result  almost coincides with a
much simpler
treatment using a mean-field expansion of the effective potential
{\it a-la} Landau  around the critical point in the limit of small quark mass.

It is perhaps premature to discuss whether a possible experimental
signature for the
QCD phase transition may be inferred from the behaviour of the  sigma
width. The mass ratio
itself is in any case a significant parameter to  describe  the degree of
chiral symmetry
breaking at different temperatures and therefore it has to be considered
as a quantity of intrinsic theoretical interest.

\acknowledgements

R. C. would like to thank Prof. J. P. Eckmann, Director of the
Department of Theoretical Physics of the University of Geneva, for
the very kind hospitality.

This work is part of the EEC project CHRXCT94/0579 (OFES 95.0200).

\end{document}